\documentclass[letterpaper,english]{achemso}
\usepackage[T1]{fontenc}
\usepackage[latin9]{inputenc}
\usepackage{amsmath}
\usepackage{amssymb}
\usepackage{graphicx}
\usepackage{esint}
\usepackage[numbers]{natbib}

\makeatletter

\title{Conformational Nonequilibrium Enzyme Kinetics: Generalized Michaelis\textendash Menten
Equation}
\author{D. Evan Piephoff}
\altaffiliation{Contributed equally to this work}
\author{Jianlan Wu}
\altaffiliation{Contributed equally to this work}
\author{Jianshu Cao}
\affiliation{Department of Chemistry, Massachusetts Institute of Technology, Cambridge,
Massachusetts 02139, United States}
\email{jianshu@mit.edu}
\pdfpageheight\paperheight
\pdfpagewidth\paperwidth

\makeatother

\usepackage{babel}
\begin{document}
\begin{abstract}
In a conformational nonequilibrium steady state (cNESS), enzyme turnover
is modulated by the underlying conformational dynamics. Based on a
discrete kinetic network model, we use the integrated probability
flux balance method to derive the cNESS turnover rate for a conformation-modulated
enzymatic reaction. The traditional Michaelis\textendash Menten (MM)
rate equation is extended to a generalized form, which includes non-MM
corrections induced by conformational population currents within combined
cyclic kinetic loops. When conformational detailed balance is satisfied,
the turnover rate reduces to the MM functional form, explaining its
validity for many enzymatic systems. For the first time, a one-to-one
correspondence is established between non-MM terms and combined cyclic
loops with unbalanced conformational currents. Cooperativity resulting
from nonequilibrium conformational dynamics has been observed in enzymatic
reactions, and we provide a novel, rigorous means of predicting and
characterizing such behavior. Our generalized MM equation affords
a systematic approach for exploring cNESS enzyme kinetics.
\end{abstract}
Conformational dynamics is essential for understanding the biological
functions of enzymes. For decades, the framework of enzymatic reactions
has been the traditional Michaelis\textendash Menten (MM) mechanism,\citep{Michaelis_1913}
where enzyme-substrate binding initializes an irreversible catalytic
reaction to form a product. The average turnover rate $v$ in a steady
state (SS) follows a hyperbolic dependence on the substrate concentration
{[}S{]}, $v=k_{2}[\mathrm{S}]/(K_{\mathrm{M}}+[\mathrm{S}])$, where
the catalytic rate $k_{2}$ and the Michaelis constant $K_{\mathrm{M}}$
characterize this enzymatic chain reaction. In contrast to the single-conformation
assumption for the traditional MM mechanism, recent single-molecule
experiments\citep{Noji_1997,English2005,Lu_2014} have revealed the
existence of multiple enzymatic conformations, spanning a broad range
of lifetime scales from milliseconds to hours. Conformational dynamics,
including hopping between different conformations and thermal fluctuations
around a single-conformation potential well, must be incorporated
into enzymatic reaction models for a quantitative study.\citep{English2005,Frieden_1979,Cornish_Bowden_1987,Cao2000,Gopich_2003,Xue_2006,Min_2008,Lomholt_2007,Xing_2007,Qian_2008,Chaudhury_2009,Cao2011,Kolomeisky_2011,Ochoa_2011,Barato_2015}
Slow conformational dynamics modulate the enzymatic reaction and allow
the enzyme to exist in a conformational nonequilibrium steady state
(cNESS), permitting complex deviations from MM kinetics (the hyperbolic
{[}S{]} dependence for \textit{v}). However, experimental and theoretical
studies have shown MM kinetics to be valid in the presence of slow
conformational dynamics under certain conditions, although $k_{2}$
and $K_{\mathrm{M}}$ become averaged over conformations.\citep{English2005,Min_2008,Cao2011}
Is there a unifying theme governing this surprising behavior?

Non-MM enzyme kinetics have been characterized by cooperativity for
many years.\citep{Frieden_1979,Cornish_Bowden_1987,Fersht_1985} For
allosteric enzymes with multiple binding sites, the binding event
at one site can alter the reaction activity at another site, accelerating
(decelerating) the turnover rate and resulting in positive (negative)
cooperativity.\citep{Fersht_1985} Another common deviation from MM
kinetics is substrate inhibition, where the turnover rate reaches
its maximum value at a finite substrate concentration and then decreases
at high substrate concentrations.\citep{Fersht_1985} For a monomeric
enzyme, the above non-MM kinetic behavior, referred to \textendash{}
in this case \textendash{} as `kinetic cooperativity,'\citep{Fersht_1985}
can be achieved by a completely different mechanism: nonequilibrium
conformational dynamics.\citep{Frieden_1979,Cornish_Bowden_1987,Lomholt_2007,Xing_2007,Qian_2008,Chaudhury_2009,Cao2011}
Can we characterize and predict this interesting behavior in a cNESS?

Recently, theoretical efforts have been applied to study conformation-modulated
enzyme kinetics by including dynamics along a conformational coordinate.
On the basis of the usual rate approach, some previous work has demonstrated
certain non-MM kinetics under specific conditions.\citep{Min_2008,Xing_2007,Qian_2008,Chaudhury_2009,Kolomeisky_2011}
Based on an alternative integrated probability flux balance method,
non-MM kinetics were linked to a nonzero conformational population
current, i.e., broken conformational detailed balance, in a two-conformation
model, and a general MM expression was speculated.\citep{Cao2011}
However, a generalized theory to systematically analyze cNESS enzyme
kinetics is still needed. In this Letter, we focus on a monomeric
enzyme and apply the integrated flux balance method to derive a generalized
form for the turnover rate, which includes non-MM corrections. We
show that when conformational detailed balance is satisfied, MM kinetics
hold, explaining their general validity. In addition, the deviations
from MM kinetics are analyzed with reduced parameters from the generalized
form of $v.$ For an extended version of our derivation, we refer
readers to ref 20.\nocite{Wu2012}

\begin{figure}
\includegraphics[width=3.25in]{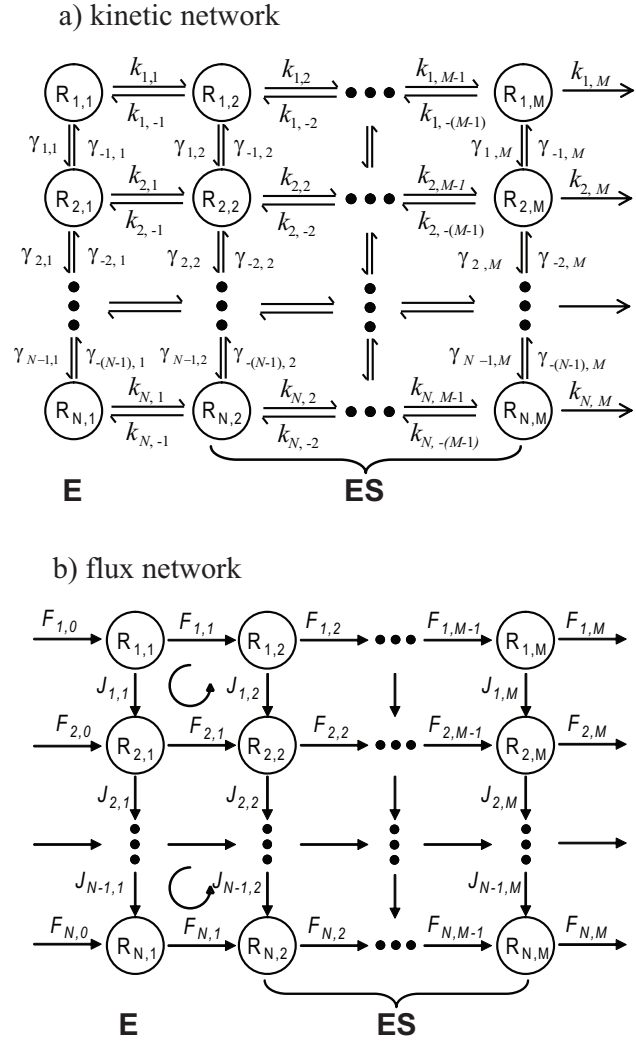}

\caption{\label{fig:kinetic_network}(a) Generalized kinetic network scheme
for a conformation-modulated enzymatic reaction. (b) Flux network
corresponding to (a) (see text for details).}
\end{figure}

To describe the generalized conformation-modulated reaction catalyzed
by a monomeric enzyme, we introduce a discrete kinetic network model,
which is illustrated in Figure \ref{fig:kinetic_network}a. This $N\times M$
network consists of a vertical conformation coordinate $(1\leq i\leq N)$
and a horizontal reaction coordinate $(1\leq j\leq M)$ . For the
reaction state index, $j=1$ denotes the initial substrate-unbound
enzymatic state (E), whereas $j\geq2$ denotes intermediate substrate-bound
enzymatic states (ES).\citep{English2005,Cao2008,Chaudhury_2009,Cao2011,Moffitt2013}
Without product states, our network corresponds to a dissipative system.
For an arbitrary site $R_{i,j}$, the reaction rates for the forward
$(R_{i,j}\rightarrow R_{i,j+1})$ and backward $(R_{i,j}\rightarrow R_{i,j-1})$
directions are given by $k_{i,j}$ and $k_{i,-(j-1)}$, respectively.
The rate for enzyme-substrate binding, the only step in our model
dependent upon substrate concentration {[}S{]}, depends linearly on
{[}S{]} as $k_{i,1}=k_{i,1}^{0}[\mathrm{S}]$ for binding rate constant
$k_{i,1}^{0}$, with {[}S{]} maintained constant in most enzymatic
experiments. The conformational dynamics are treated via a kinetic
rate approach, with the interconversion (hopping or diffusion) rates
for $R_{i,j}\rightarrow R_{i+1,j}$ and $R_{i,j}\rightarrow R_{i-1,j}$
given by $\gamma_{i,j}$ and $\gamma_{-(i-1),j}$, respectively. We
note that local detailed balance results in the constraint $k_{i,j}\gamma_{-i,j}/(k_{i,-j}\gamma_{i,j})=k_{i+1,j}\gamma_{-i,j+1}/(k_{i+1,-j}\gamma_{i,j+1})$
for $j\leq M-1$. However, for the purposes of our kinetic analysis,
it is unnecessary to impose this relation, as our principal results
hold, irrespective of whether it is satisfied. The rate equation for
site $R_{i,j}$ is written as
\begin{equation}
\frac{d}{dt}P_{i,j}(t)=\sum_{i'=1}^{N}\gamma_{i,i';j}P_{i',j}(t)+\sum_{j'=1}^{M}k_{j,j';i}P_{i,j'}(t)\label{eq:rate_eqn}
\end{equation}
where $P_{i,j}(t)$ is the probability of an enzyme in site $R_{i,j}$
at time $t$, i.e., the survival probability for the site. Here, $\gamma_{i,i';j}=\gamma_{i-1,j}\delta_{i',i-1}+\gamma_{-i,j}\delta_{i',i+1}-[\gamma_{i,j}+\gamma_{-(i-1),j}]\delta_{i',i}$
denotes the interconversion rates in the $j$-th reaction state and
$k_{j,j';i}=k_{i,j-1}\delta_{j',j-1}+\gamma_{i,-j}\delta_{j',j+1}-[k_{i,j}+k_{i,-(j-1)}]\delta_{j',j}$
denotes the reaction rates for the $i$-th conformation.

Within the framework of a dissipative enzymatic network, the average
turnover rate $v$ is equivalent to the inverse of the mean first
passage time (MFPT) $\langle t\rangle$. Using the residence time
${\displaystyle \tau_{i,j}=\int_{0}^{\infty}P_{i,j}(t)dt}$ at each
site $R_{i,j}$, we can express the MFPT in the $N\times M$ network
as a summation of $\tau_{i,j}$, i.e, $\langle t\rangle=\sum_{i,j}\tau_{i,j}$.
Instead of inverting the transition matrix,\citep{English2005} we
evaluate $\tau_{i,j}$ by inspecting integrated probability fluxes,\citep{Cao2011}
which correspond to stationary population fluxes normalized by $v$,
and these will be shown to directly reflect conformational nonequilibrium.
Along the horizontal reaction coordinate, the integrated flux for
$R_{i,j}\rightarrow R_{i,j+1}$ is given by $F_{i,j}=k_{i,j}\tau_{i,j}-k_{i,-j}\tau_{i,j+1}$.
Along the vertical conformation coordinate, the integrated flux for
$R_{i,j}\rightarrow R_{i+1,j}$ is given by $J_{i,j}=\gamma_{i,j}\tau_{i,j}-\gamma_{-i,j}\tau_{i+1,j}$.
In addition, we need to specify the initial condition $P_{i,j}(t=0)$
for calculating $\langle t\rangle$. For a monomeric enzyme, each
turnover event begins with the substrate-unbound state, and $P_{i,1}(t=0)$
defines the initial flux $F_{i,0}.$ With the definition of $\{F_{i,j},\ J_{i,j}\}$,
we map the original kinetic network to a flux network as shown in
Figure \ref{fig:kinetic_network}b. For each site $R_{i,j}$, the
rate equation in eq \ref{eq:rate_eqn} is replaced by a flux balance
relation,
\begin{equation}
F_{i,j-1}+J_{i-1,j}=F_{i,j}+J_{i,j}\label{eq:flux_bal_rel}
\end{equation}
which is generalized to the probability conservation law: \textit{the
total input integrated probability flux must equal the total output
integrated probability flux}. This conservation law can be extended
to complex first-order kinetic structures including the $N\times M$
network. The flux balance method thus provides a simple means of calculating
the MFPT.

To evaluate the MFPT, we begin with the final reaction state $(j=M)$
and propagate all the fluxes back to the initial reaction state $(j=1)$
based on eq \ref{eq:flux_bal_rel}. For each site $R_{i,j}$, the
physical nature of the first-order kinetics determines that all three
variables, $\tau_{i,j}$, $J_{i,j}$ and $F_{i,j}$, are linear combinations
of terminal fluxes $F_{i,j=M}$. The first two variables are formally
written as $\tau_{i,j}=\sum_{i'=1}^{N}a_{i,j,i'}F_{i',M}$ and $J_{i,j}=\sum_{i'}{\displaystyle c_{i,j,i'}F_{i',M}}$,
where $a_{i,j,i'}$ and $c_{i,j,i'}$ are coefficients depending on
rate constants $\{k,\ \gamma\}$. For example, the coefficients for
the final reaction state are $a_{i,M,i'}=1/k_{i,M}\delta_{i',i}$
and $c_{i,M,i'}=\gamma_{i,M}/k_{i,M}\delta_{i',i}-\gamma_{-i,M}/k_{i+1,M}\delta_{i',i+1}$.
Because of the direction of our reversed flux propagation, only the
coefficients for the initial reaction state are {[}S{]} dependent,
and they can be explicitly written as $a_{i,1,i'}=b_{i,i'}/[\mathrm{S}]$
and $c_{i,1,i'}=d_{i,i'}/[\mathrm{S}]$. The substrate-unbound $(\mathrm{E}_{i}=R_{i,1})$
and substrate-bound ($\mathrm{E}\mathrm{S}_{i}=\sum_{j=2}^{M}R_{i,j}$)
states are distinguished by the different $[\mathrm{S}]$ dependence
of the coefficients. The MFPT is thus given by
\begin{equation}
\langle t\rangle={\displaystyle \sum_{i'=1}^{N}\left[\frac{\sum_{i=1}^{N}b_{i,i'}}{[\mathrm{S}]}+\sum_{i=1}^{N}\sum_{j=2}^{M}a_{i,j,i'}\right]F_{i',M}}\label{eq:mfpt}
\end{equation}

The essential part of our derivation is then to solve for the terminal
fluxes $F_{i,M}$. The SS condition can be interpreted as follows:
after each product release, the enzyme returns to the same conformation
for the next turnover reaction, i.e., $F_{i,M}=F_{i,0}$.\citep{Cao2000}
Applying the probability conservation law to each horizontal chain
reaction with a single conformation and considering the boundary condition
at conformations $i=1$ and $N$, we express the SS condition as a
flux constraint, $J_{i,\mathrm{E}}+J_{i,\mathrm{E}\mathrm{S}}=0$
for $i=1,2,\cdots,N-1$, where $J_{i,\mathrm{E}}=J_{i,1}$ and $J_{i,\mathrm{E}\mathrm{S}}=\sum_{j=2}^{M}J_{i,j}$.
For each combined cyclic loop $\mathrm{E}_{i}\rightarrow\mathrm{E}_{i+1}\rightarrow\mathrm{E}\mathrm{S}_{i+1}\rightarrow\mathrm{E}\mathrm{S}_{i}\rightarrow\mathrm{E}_{i}$,
there may exist a stabilized nonequilibrium conformational population
current (see Figure \ref{fig:kinetic_network}b), with $J_{i,\mathrm{E}}$
representing this stationary current normalized by $v$. However,
under certain circumstances, $J_{i,\mathrm{E}}$ can vanish, and the
SS condition is further simplified to $J_{i,\mathrm{E}\mathrm{S}}=0$.
We note that satisfaction of the aforementioned constraint resulting
from local detailed balance still permits nonzero $J_{i,\mathrm{E}}$.
In general, we assume that there exist $N_{c}(\leq N-1)$ nonzero
conformational currents and $(N-1-N_{c})$ zero ones. In addition
to these $(N-1)$ current conditions, the normalization condition
$\sum_{i=1}^{N}F_{i,0}=1$ is needed for fully determining the initial
fluxes (due to $F_{i,0}=F_{i,M}$). As a result, we derive an $N$-equation
array for $F_{i,0},$
\begin{align}
 & \mathbf{U}\cdot\mathbf{F}=\nonumber \\
 & \begin{bmatrix}1 & 1 & \cdots\\
C_{1,1}+\left(\frac{d_{1,1}}{\left[\mathrm{S}\right]}\right) & C_{1,2}+\left(\frac{d_{1,2}}{\left[\mathrm{S}\right]}\right) & \cdots\\
C_{2,1}+\left(\frac{d_{2,1}}{\left[\mathrm{S}\right]}\right) & C_{2,2}+\left(\frac{d_{2,2}}{\left[\mathrm{S}\right]}\right) & \cdots\\
\vdots & \vdots & \vdots
\end{bmatrix}\cdot\begin{bmatrix}F_{1,0}\\
F_{2,0}\\
\vdots
\end{bmatrix}=\begin{bmatrix}1\\
0\\
\vdots
\end{bmatrix}\label{eq:eq_array}
\end{align}
with $C_{i,i'}=\sum_{j=2}^{M}c_{i,j,i'}$. Notice that for the $(i+1)$-th
row of matrix $\mathbf{U}$ in eq \ref{eq:eq_array}, $d_{i,i'}/[\mathrm{S}]$
only exists when $J_{i,\mathrm{E}}\neq0$, and $N_{c}$ rows are {[}S{]}
dependent for this matrix. We solve for the initial fluxes by the
matrix inversion $F_{i,0}=[\mathrm{\mathbf{U}}^{-1}]_{i,1}$. After
a tedious but straightforward derivation, $F_{i,0}$ is written as
\begin{equation}
F_{i,0}([{\displaystyle \mathrm{S}])=f_{i,0}+\sum_{n=1}^{N_{c}}f_{i,n}/([\mathrm{S}]+s_{n})}\label{eq:initial_flux}
\end{equation}
where each $s_{n}$ is assumed to be distinct, and constraints hold
for $\sum_{i}f_{i,0}=1$ and $\sum_{i}f_{i,n}=0$ for $n\geq1$.

Substituting eq \ref{eq:initial_flux} into eq \ref{eq:mfpt}, we
obtain the key result of this Letter: the cNESS turnover rate for
the $N\times M$ network with $N_{c}$ unbalanced conformational currents
is given by a generalized Michaelis\textendash Menten equation,
\begin{equation}
v=\left[A_{0}+{\displaystyle \frac{B_{0}}{[\mathrm{S}]}+\sum_{n=1}^{N_{c}}\frac{B_{n}}{[\mathrm{S}]+s_{n}}}\right]^{-1}\label{eq:turnover_rate}
\end{equation}
where the reduced parameters are $A_{0}=\langle1/k_{2}^{\mathrm{e}\mathrm{f}\mathrm{f}}\rangle_{[\mathrm{S}]\rightarrow\infty}$,
$B_{0}=\langle K_{\mathrm{M}}^{\mathrm{e}\mathrm{f}\mathrm{f}}/k_{2}^{\mathrm{e}\mathrm{f}\mathrm{f}}\rangle_{[\mathrm{S}]=0}$,
and $B_{n}=\sum_{i}[1/k_{i,2}^{\mathrm{e}\mathrm{f}\mathrm{f}}-K_{i,\mathrm{M}}^{\mathrm{e}\mathrm{f}\mathrm{f}}/(k_{i,2}^{\mathrm{e}\mathrm{f}\mathrm{f}}s_{n})]f_{i,n}$.
For each conformational channel, we introduce an effective catalytic
rate $k_{i,2}^{\mathrm{e}\mathrm{f}\mathrm{f}}=(\sum_{i'=1}^{N}\sum_{j=2}^{M}a_{i',j,i})^{-1}$
and an effective Michaelis constant $K_{i,\mathrm{M}}^{\mathrm{e}\mathrm{f}\mathrm{f}}=k_{i,2}^{\mathrm{e}\mathrm{f}\mathrm{f}}\sum_{i'}b_{i',i}$,
which describe the kinetics within that channel in the decomposed
representation of the scheme, wherein the $N$ two-state chain reactions
are effectively independent, each with probability $F_{i,0}$. The
conformational average is defined as $\langle x\rangle_{[\mathrm{S}]}=\sum_{i}x_{i}F_{i,0}([\mathrm{S}])$
for a conformation-dependent variable $x_{i}$. In the right hand
side of eq \ref{eq:turnover_rate}, the first two terms retain the
traditional MM form, whereas the remaining $N_{c}$ terms introduce
non-MM rate behavior, with a 1:1 correspondence between non-MM terms
and combined cyclic loops with nonzero conformational currents. Our
derivations clearly show that these non-MM terms are induced by the
{[}S{]}-dependent conformational distribution $\mathbf{F}$ resulting
from nonequilibrium conformational currents. Therefore, MM kinetics
are valid when conformational detailed balance is satisfied, where
all $B_{n}$ vanish due to $F_{i,0}=f_{i,0}.$

\begin{figure}
\includegraphics[width=3.25in]{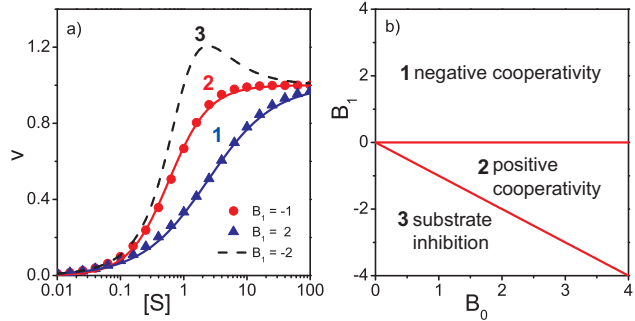}

\caption{\label{fig:single_loop}(a) Three non-MM turnover rates for the single-loop
model with $A_{0}=B_{0}=s_{1}=1$. The circles $(B_{1}=-1)$ and the
up-triangles $(B_{1}=2)$ exhibit positive and negative cooperativity,
respectively. The two solid lines are the fit using the Hill equation.
The dashed line $(B_{1}=-2)$ shows substrate inhibition behavior.
(b) Phase diagram of enzyme kinetics for the single-loop model. Two
lines, $B_{1}=0$ and $B_{1}=-B_{0}$, separate three regimes of kinetics.}
\end{figure}

With nonzero conformational currents, the enzyme kinetics are expected
to exhibit cooperative non-MM behavior. As a demonstration, the single-loop
model with only one current $J_{1,\mathrm{E}}$ and one non-MM term
$B_{1}/([\mathrm{S}]+s_{1})$ is first considered. With other parameters
fixed, we calculate turnover rates $v$ for the three values of $B_{1}$
in Figure \ref{fig:single_loop}a. For the two turnover rates monotonically
increasing with {[}S{]} ($B_{1}=-1$ and $2$), we fit them with the
Hill equation, $v/v_{\max}=[\mathrm{S}]^{n_{\mathrm{H}}}/(\kappa+[\mathrm{S}]^{n_{\mathrm{H}}})$,
where the Hill constant $n_{\mathrm{H}}>1$ $(n_{\mathrm{H}}<1)$
indicates positive (negative) cooperativity. The fitting results show
that cooperativity is completely determined by the sign of $B_{1}$:
positive for $B_{1}<0$ and negative for $B_{1}>0$. This result is
also reflected in eq \ref{eq:turnover_rate}, where negative (positive)
$B_{1}$ increases (decreases) the MM turnover rate $(A_{0}+B_{0}/[\mathrm{S}])^{-1}$.
The dashed line in Figure \ref{fig:single_loop}a shows that a largely
negative $B_{1}+B_{0}$ leads to substrate inhibition behavior. The
cNESS substrate inhibition shows positive cooperativity at low substrate
concentrations, and then the turnover rate decreases to a nonzero
value $A_{0}^{-1}$ in the substrate-saturation limit. Next, we plot
the phase diagram of enzyme kinetics for the single-loop model in
Figure \ref{fig:single_loop}b, which only depends on $B_{0}$ and
$B_{1}.$ From this phase diagram, $\alpha=B_{1}/B_{0}$ is defined
as a unique non-MM indicator for single-loop systems, with negative
cooperativity for $\alpha>0$, positive cooperativity for $-1\leq\alpha<0$,
and substrate inhibition for $\alpha<-1.$

\begin{figure}
\includegraphics[width=3.25in]{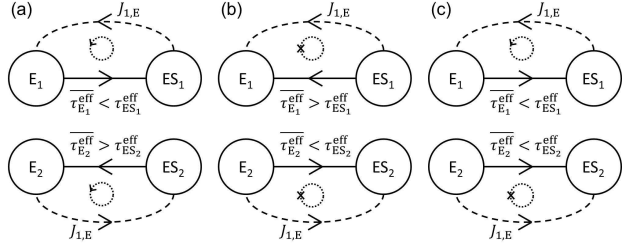}

\caption{\label{fig:current_mod}(a)\textendash (c) Three cases in which a
current $J_{1,\mathrm{E}}$ circulating counterclockwise within a
two-conformation loop can be modulated by $\Delta\overline{\Delta\tau^{\mathrm{eff}}}$
(see text for details); such modulation underlies the emergence of
kinetic cooperativity. In each conformational channel, a horizontal
arrow proceeds from the state with the faster effective characteristic
residence time (see text for details) to the state with the slower
one, with $J_{1,\mathrm{E}}$ superimposed onto this view. Note that
there are also analogous cases for $J_{1,\mathrm{E}}$ proceeding
in the clockwise direction.}
\end{figure}

The direction of a conformational current alone does not predict its
influence on the cooperativity, which raises the question of how currents
are modulated to govern cooperative behavior. For the two-conformation
network, the simplest single-loop model, we can rewrite the non-MM
term as
\begin{equation}
\frac{B_{1}}{[\mathrm{S}]+s_{1}}\propto\Delta\overline{\Delta\tau^{\mathrm{eff}}}\times J_{1,\mathrm{E}}([\mathrm{S}])
\end{equation}
where $\Delta\overline{\Delta\tau^{\mathrm{eff}}}=\overline{\Delta\tau^{\mathrm{eff}}}_{1}-\overline{\Delta\tau^{\mathrm{eff}}}_{2}$,
with $\overline{\Delta\tau^{\mathrm{eff}}}_{i}=\overline{\tau_{\mathrm{E}_{i}}^{\mathrm{eff}}}-\tau_{\mathrm{ES}_{i}}^{\mathrm{eff}}$.
Here, the $\mathrm{E}_{i}$ and $\mathrm{ES}_{i}$ residence times
in the decomposed representation, each independent of the non-MM term
{[}and thus of $J_{1,\mathrm{E}}([\mathrm{S}])${]}, are given by
$\tau_{\mathrm{E}_{i}}^{\mathrm{eff}}([\mathrm{S}])=K_{i,\mathrm{M}}^{\mathrm{e}\mathrm{f}\mathrm{f}}/(k_{i,2}^{\mathrm{e}\mathrm{f}\mathrm{f}}[\mathrm{S}])$
and $\tau_{\mathrm{ES}_{i}}^{\mathrm{eff}}=1/k_{i,2}^{\mathrm{e}\mathrm{f}\mathrm{f}}$,
respectively. Also, $\overline{\tau_{\mathrm{E}_{i}}^{\mathrm{eff}}}=\tau_{\mathrm{E}_{i}}^{\mathrm{eff}}([\mathrm{S}]=s_{1})$,
where $s_{1}$ is the value of $[\mathrm{S}]$ at which $|J_{1,\mathrm{E}}([\mathrm{S}])|$
is at half its maximum and thus represents a characteristic non-MM
substrate concentration. Therefore, $\overline{\tau_{\mathrm{E}_{i}}^{\mathrm{eff}}}$
represents a characteristic value of $\tau_{\mathrm{E}_{i}}^{\mathrm{eff}}([\mathrm{S}])$,
with corresponding characteristic residence time gradient $\overline{\Delta\tau^{\mathrm{eff}}}_{i}$.
Thus, $\Delta\overline{\Delta\tau^{\mathrm{eff}}}$ represents the
difference in characteristic residence time gradient between the two
decomposed conformational channels. Cooperativity depends upon $J_{1,\mathrm{E}}$
modulated by $\Delta\overline{\Delta\tau^{\mathrm{eff}}}$, i.e.,
it is governed by the relative modulation of the current between the
two decomposed chain reactions. In the two-conformation model, $J_{1,\mathrm{E}}$
proceeds from $\mathrm{E}_{i}$ to $\mathrm{ES}_{i}$ in one conformational
channel and from $\mathrm{ES}_{i}$ to $\mathrm{E}_{i}$ in the other,
as illustrated in Figure \ref{fig:current_mod}a\textendash c for
a counterclockwise current, which corresponds to $J_{1,\mathrm{E}}>0$
based upon our original definition of $J_{i,j}$. In each two-state
chain reaction, enzyme turnover is accelerated (decelerated) when
$J_{1,\mathrm{E}}$ proceeds from the state with the slower (faster)
effective characteristic residence time to the state with the faster
(slower) one. In Figure \ref{fig:current_mod}a (b), turnover is accelerated
(decelerated) in both chain reactions, resulting in overall turnover
acceleration (deceleration), i.e., positive cooperativity or substrate
inhibition (negative cooperativity). In Figure \ref{fig:current_mod}c,
turnover is accelerated in conformation 1 and decelerated in conformation
2 (the opposite {[}not shown{]} is possible as well), with the cooperativity
depending upon the relative modulation of the current between the
two decomposed chains. Kinetic cooperativity is thus explained as
follows: when $J_{1,\mathrm{E}}$ proceeds in the direction that,
on average, corresponds to decreasing (increasing) effective characteristic
residence time, positive cooperativity or substrate inhibition (negative
cooperativity) occurs.

Interestingly, when the effective characteristic residence time gradient
is conformation invariant, modulation of the conformational current
is balanced, resulting in MM kinetics, even in the presence of circulating
current (i.e., when $\Delta\overline{\Delta\tau^{\mathrm{eff}}}=0$
and $J_{1,\mathrm{E}}\neq0$ for the two-conformation network). This
scenario represents a unique type of nonequilibrium symmetry in multidimensional
kinetic networks and is not precluded by the satisfaction of the aforementioned
constraint resulting from local detailed balance. Additionally, we
note that for the $2\times2$ model, $J_{1,\mathrm{E}}$ vanishes
under a simple conformational detailed balance condition,
\begin{equation}
\frac{\gamma_{1,1}}{\gamma_{1,2}}K_{1,\mathrm{M}}=\frac{\gamma_{-1,1}}{\gamma_{-1,2}}K_{2,\mathrm{M}}
\end{equation}
where $K_{i,\mathrm{M}}=(k_{i,-1}+k_{i,2})/k_{i,1}^{0}$. Explicit
calculations for this model are provided in the Supporting Information.

\begin{figure}
\includegraphics[width=3.25in]{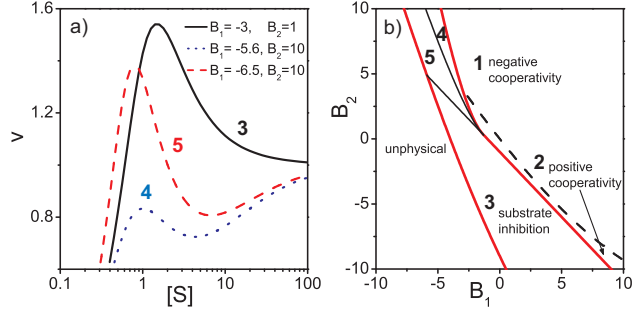}

\caption{\label{fig:two_loop}Enzyme kinetics for the two-loop model with $A_{0}=B_{0}=s_{1}=1$
and $s_{2}=4$. (a) Three turnover rates $v$ that are non-monotonic
functions of {[}S{]}. Each line shows a typical type of non-MM kinetic
behavior from the regime labeled by the same number in (b). (b) Phase
diagram determined by two non-MM parameters $B_{1}$ and $B_{2}$.
There are five regimes of non-MM behavior (see text for details).}
\end{figure}

For the two-loop models with two non-MM terms, the cNESS enzyme kinetics
become more complicated, as illustrated in a typical phase diagram
in Figure \ref{fig:two_loop}b. Except for an unphysical regime where
$v$ shows divergence and negativity, five regimes of enzyme kinetics
can be characterized in the phase space composed of $B_{1}$ and $B_{2}$.
Similar to the single-loop model, when monotonically increasing to
the maximum value $v_{\max}$ in the substrate-saturation limit $([\mathrm{S}]\rightarrow\infty)$,
$v$ can exhibit negative (Regime 1) and positive (Regime 2) cooperativity.
The separation line between these two kinetic regimes, however, is
hard to rigorously define. The dashed separation line in Figure \ref{fig:two_loop}b
corresponds to $n_{\mathrm{H}}=1$, where the Hill constant is empirically
calculated using $n_{\mathrm{H}}=\log81/\log([\mathrm{S}]_{0.9v_{\max}}/[\mathrm{S}]_{0.1v_{\max}})$,\citep{Fersht_1985}
and $[\mathrm{S}]_{v}$ is the substrate concentration for $v$. In
Regimes 3-5, the turnover rate $v$ is a non-monotonic function of
{[}S{]} (examples shown in Figure \ref{fig:two_loop}a). In Regime
3, with $v_{\max}$ occurring at a finite $[\mathrm{S}]_{v_{\max}}$,
the turnover rate exhibits the same substrate inhibition behavior
as the single-loop model. Alternatively, an additional local minimum
of $v$ can appear at $[\mathrm{S}]_{v_{\mathrm{min}}}(>[\mathrm{S}]_{v_{\max}})$,
and $v$ increases at high substrate concentrations instead. Two examples
are shown by the dashed and dotted lines in Figure \ref{fig:two_loop}a.
Based on a criterion whether the global $v_{\max}$ appears as $\left[\mathrm{S}\right]\rightarrow\infty$
or at the finite $[\mathrm{S}]_{v_{\max}}$, this non-MM kinetic behavior
is further divided into Regimes 4 and 5, respectively.

For the generalized $N_{c}$-loop model, cNESS enzyme kinetics can
be similarly analyzed using reduced parameters from eq \ref{eq:turnover_rate}.
In the case that all the non-MM parameters $B_{n}$ are positive (negative),
the turnover rate exhibits negative cooperativity (positive cooperativity
or substrate inhibition). With the coexistence of positive and negative
non-MM parameters, cooperativity can be qualitatively determined by
the small-$[\mathrm{S}]$ expansion of the turnover rate in eq \ref{eq:turnover_rate},
$v\sim B_{0}^{-1}[\mathrm{S}]-(A_{0}B_{0}^{-2}+\sum_{n=1}^{N_{c}}B_{n}s_{n}^{-1}B_{0}^{-1})[\mathrm{S}]^{2}+O([\mathrm{S}]^{3})$.
For a largely negative $\sum_{n}B_{n}/s_{n}$, the positive quadratic
{[}S{]} term dominates in $v$, resulting in positive cooperativity.
When this summation becomes largely positive, the cancellation between
linear and nonlinear terms can slow down the increase of $v$ with
{[}S{]}, inducing negative cooperativity. The sign of $\sum_{n}B_{n}/s_{n}$
is thus a qualitative indicator of cooperativity. To investigate the
substrate inhibition behavior, we expand $v$ in the substrate-saturation
limit as $v\sim A_{0}^{-1}-(B_{0}+\sum_{n}B_{n})A_{0}^{-2}[\mathrm{S}]^{-1}+O([\mathrm{S}]^{-2})$.
For $\sum_{n}B_{n}<-B_{0}$, ${\displaystyle v}$ is a decreasing
function of {[}S{]}, and the maximum turnover rate $v_{\max}$ must
appear at a finite {[}S{]}. The investigation of other types of non-monotonic
behavior for $v$ needs the explicit rate form in eq \ref{eq:turnover_rate}.

In summary, we study cNESS enzyme kinetics induced by population currents
from conformational dynamics. Applying the flux balance method to
a discrete $N\times M$ kinetic model, we derive a generalized Michaelis\textendash Menten
equation to predict the {[}S{]} dependence of the turnover rate. Using
reduced non-MM parameters, $B_{n}$ in eq \ref{eq:turnover_rate},
our generalized MM equation provides a systematic approach to explore
cNESS enzyme kinetics. Compared to the typical rate matrix approach,
our flux method characterizes non-MM enzyme kinetics in a much simpler
way. For example, a unique kinetic indicator $\alpha=B_{1}/B_{0}$
is defined for the single-loop model, and phase diagrams are plotted
for the single- and two-loop models. Our study can be extended to
other important biophysical processes following the MM mechanism,
e.g., the movement of molecular motors induced by ATP binding.

\section*{Acknowledgments}

This work was supported by the NSF (Grant No. CHE-1112825) and the
Singapore-MIT Alliance for Research and Technology (SMART). D. E.
P. acknowledges support from the NSF Graduate Research Fellowship
Program.

\section*{Supporting Information Available}

Four-site, single-loop model calculations

\bibliographystyle{achemso}
\providecommand{\latin}[1]{#1}
\providecommand*\mcitethebibliography{\thebibliography}
\csname @ifundefined\endcsname{endmcitethebibliography}
  {\let\endmcitethebibliography\endthebibliography}{}

\end{document}